\documentclass[aps,prx,reprint,superscriptaddress,groupedaddress]{revtex4-2}
\usepackage{amsmath}
\usepackage{amssymb}
\usepackage{physics}
\usepackage{bm}

\begin{document}

\title{Coherent Quantum Schr\"odinger Bridge: 
\\
Two-Boundary Optimal Control for Quantum Algorithm Design}

\author{Masayuki Ohzeki}
\email{mohzeki@tohoku.ac.jp}
\affiliation{Graduate School of Information Sciences, Tohoku University, Sendai 980-8579, Japan}
\affiliation{Department of Physics, Institute of Science, Tokyo 162-8601, Japan}
\affiliation{Research and Education Institute for Semiconductors and Informatics, Kumamoto University, Kumamoto 860-8555, Japan}
\affiliation{Sigma-i Co., Ltd., Tokyo 108-0075, Japan}

\date{\today}

\begin{abstract}
Quantum algorithms are intrinsically two-boundary processes: an input state is prepared, and an output state or subspace is selected as the computational answer.
We formulate this observation as a coherent Quantum Schr\"odinger Bridge (QSB), a pure-state Hamiltonian counterpart of Schr\"odinger bridge theory in which the endpoint constraint is imposed on state vectors and the transport cost is the quadratic control action.
In this setting Aharonov's two-state vector becomes the natural optimal-control pair: a forward state from the input and a backward state from the target.
Pontryagin's principle then yields a universal optimal Hamiltonian whose weak value is purely imaginary in the geodesic gauge.
Thus weak values are not an auxiliary interpretation; they are the local response functions that quantify the drift of the pre-selected state toward the post-selected boundary.
Applying this framework to unstructured search, periodicity finding, and matrix arithmetic, we reconstruct Grover's algorithm, the quantum Fourier transform underlying Shor's algorithm, and quantum singular value transformation (QSVT).
The usual circuit components---oracles, diffusion reflections, controlled phases, and signal-processing rotations---emerge as Lie-algebraic syntheses of the optimal weak-value drift.
This perspective unifies distinct algorithmic paradigms into a single geometric principle: algorithm design is the problem of choosing computational boundary conditions and realizing the corresponding optimal flow.
\end{abstract}

\maketitle

\section{Introduction}
The landscape of quantum algorithm design has traditionally been fragmented into distinct paradigms. 
Seminal algorithms such as Grover's unstructured search \cite{Grover1997}, Shor's integer factorization \cite{Shor1994}, and the more recent Quantum Singular Value Transformation (QSVT) \cite{Gilyen2019} are often constructed using disparate heuristic gadgets---amplification, phase kickback, and signal processing---making the underlying unifying principle difficult to discern.
While the Quantum Brachistochrone formulation \cite{Carlini2006} provided a geometric perspective by characterizing algorithms as time-optimal geodesics, the constructive link between these abstract paths and the concrete circuit components---oracles, diffusion operators, Fourier transforms, and signal-processing rotations---remains under-explored.

The central point of this work is that a quantum algorithm is not merely an initial-value problem.
It is specified by what is prepared and by what must be obtained.
This two-boundary character is precisely the structure emphasized in time-symmetric quantum mechanics: between a pre-selection and a post-selection, the system is described by a forward-evolving state and a backward-evolving state \cite{Aharonov1964,AharonovVaidman1991}.
The language of weak measurement then becomes natural rather than decorative, because weak values are the local response functions of such pre- and post-selected ensembles \cite{Aharonov1988,Dressel2014}.

We turn this observation into a constructive principle by formulating quantum algorithm design as a coherent Quantum Schr\"odinger Bridge (QSB).
The word ``bridge" is used here in a deliberately restricted sense.
Schr\"odinger's original problem concerns the most likely stochastic process connecting two probability distributions \cite{Schrodinger1931}, and modern quantum variants formulate analogous endpoint problems for density matrices, quantum channels, or quantum Markov dynamics \cite{GeorgiouPavon2015,BeckerLi2021,Movilla2025}.
The present work does not claim to solve that full non-commutative stochastic bridge problem.
Instead, it takes the coherent pure-state limit relevant for circuit algorithms: the endpoints are rank-one computational boundary states, the admissible dynamics are unitary Hamiltonian flows, and the bridge cost is the quadratic control action.
In Hilbert space this becomes an energy-optimal control problem: given an input boundary and an output boundary, find the Hamiltonian of minimum control energy that transports one to the other.
The resulting optimal Hamiltonian is governed by the imaginary part of a weak value.
It therefore describes not just a kinematic rotation, but a directed drift that increases the probability amplitude of the post-selected computational answer.

We demonstrate this principle on three algorithmic primitives.
For Grover search, the oracle and diffusion reflections synthesize the optimal drift between the uniform state and the marked state.
For the quantum Fourier transform (QFT), the controlled-phase gates arise as the pairwise interactions required to imprint the target phase correlations.
For QSVT, the alternating block encoding and phase rotations synthesize singular-value-dependent optimal rotations in each invariant two-dimensional subspace.
These examples support a single message: standard quantum circuits are discrete realizations of optimal two-boundary flows generated by the non-commutativity of the available controls and the target boundary conditions.

\section{Coherent Quantum Schr\"odinger Bridge}
We consider a quantum system in an $N$-dimensional Hilbert space $\mathcal{H}$. 
The goal is to steer an initial state $\ket{\psi_{in}}$ to a target state $\ket{\psi_{target}}$ within a finite time $T$, minimizing the control energy.
Let $\ket{\psi(t)}$ be the state vector governed by the Schr\"odinger equation $(\hbar=1)$:
\begin{equation}
    \frac{d}{dt}\ket{\psi(t)} = -i H(t) \ket{\psi(t)},
\end{equation}
where $H(t)$ is the control Hamiltonian. 
We define the cost function $J[H]$ as the integrated Frobenius norm of the Hamiltonian (representing the control effort or kinetic energy on the manifold) \cite{Peirce1988, Nielsen2006}:
\begin{equation}
    J[H] = \frac{1}{2} \int_{0}^{T} \Tr(H(t)^2) \, dt.
\end{equation}
The problem is to find $H(t)$ that minimizes $J[H]$ subject to boundary conditions $\ket{\psi(0)} = \ket{\psi_{in}}$ and $\ket{\psi(T)} = e^{i\theta}\ket{\psi_{target}}$, where $\theta$ is an arbitrary global phase.
The arbitrary phase is fixed below by the geodesic gauge, in which the instantaneous overlap of the forward and backward states is real and positive.

We apply Pontryagin's minimum principle \cite{Pontryagin1962}, adapted for quantum systems \cite{Peirce1988, Khaneja2001}, by introducing the adjoint state $\ket{\chi(t)}$ as the generalized momentum conjugate to $\ket{\psi(t)}$.
The terminal condition fixes $\ket{\chi(T)} \propto \ket{\psi_{target}}$; equivalently, $\ket{\chi(t)}$ is the backward-evolving post-selected state of the two-state vector formalism.
With a sign convention chosen so that the optimal velocity points toward the post-selected boundary, the Pontryagin Hamiltonian is
\begin{equation}
    \mathcal{K} = \frac{1}{2}\Tr(H^2) - \Re \mel{\chi(t)}{-i H(t)}{\psi(t)}.
\end{equation}

The optimality condition requires $\partial \mathcal{K} / \partial H = 0$. Considering a variation $\delta H$:
\begin{align}
    \delta \mathcal{K} &= \Tr(H \delta H) - \Im \mel{\chi}{\delta H}{\psi} \nonumber \\
    &= \Tr \left[ \left( H - \frac{1}{2i} (\ketbra{\psi}{\chi} - \ketbra{\chi}{\psi}) \right) \delta H \right] = 0.
\end{align}
Solving for $H$, we obtain the optimal control Hamiltonian:
\begin{equation}
    H^*(t) = \frac{i}{2} \left( \ketbra{\chi(t)}{\psi(t)} - \ketbra{\psi(t)}{\chi(t)} \right).
    \label{eq:optimal_H}
\end{equation}
The adjoint state satisfies the corresponding canonical equation,
\begin{equation}
    \frac{d}{dt}\ket{\chi(t)} = -i H^*(t) \ket{\chi(t)}.
\end{equation}

Equation (\ref{eq:optimal_H}) is the point where optimal control, two-state quantum mechanics, and weak measurement meet.
At an intermediate time, let a small additional unitary $e^{-igA}$ be applied between the pre-selected state $\ket{\psi(t)}$ and the post-selected state $\ket{\chi(t)}$.
The post-selection probability is
\begin{equation}
    P_A(g)=\left|\mel{\chi(t)}{e^{-igA}}{\psi(t)}\right|^2.
\end{equation}
Its first-order response is
\begin{equation}
    \left.\frac{d}{dg}\ln P_A(g)\right|_{g=0}
    = 2 \Im A_w,\qquad
    A_w \equiv \frac{\mel{\chi}{A}{\psi}}{\braket{\chi}{\psi}} .
    \label{eq:weak_response}
\end{equation}
Thus the imaginary part of the weak value is the local drift of the post-selection probability, while the real part describes phase accumulation \cite{Aharonov1988,Dressel2014}.
Weak measurement is therefore not a side remark in the present framework: it supplies the differential diagnostic that tells which generator most efficiently moves the input boundary toward the output boundary.

Substituting our optimal Hamiltonian $H^*$ from Eq.(\ref{eq:optimal_H}) into the definition of the weak value, we obtain:
\begin{equation}
    H_w^* = \frac{i}{2} \left( \frac{1}{\braket{\chi}{\psi}} - \braket{\chi}{\psi} \right).
    \label{eq:imaginary_weak_value}
\end{equation}
In the geodesic gauge this weak value is purely imaginary.
A non-zero real part would spend control energy on phase accumulation rather than on increasing the post-selection probability.
The QSB solution removes this waste: the whole available generator is aligned with the imaginary weak-value response, giving the steepest fidelity growth compatible with the energy metric.
In this precise sense the optimal quantum algorithm is a weak-value drift field generated by the chosen final boundary.

The action of $H^*$ on the state can be rewritten as:
\begin{equation}
    H^* \ket{\psi} = \frac{i}{2} \braket{\chi}{\psi} \left( \ket{\psi_w} - \ket{\psi} \right),
    \qquad
    \ket{\psi_w} \equiv \frac{\ket{\chi}}{\braket{\chi}{\psi}} .
\end{equation}
Equivalently,
\begin{equation}
    \frac{d}{dt}\ket{\psi}
    = -iH^*\ket{\psi}
    = \frac{1}{2}\left(\ket{\chi}-\braket{\chi}{\psi}\ket{\psi}\right).
    \label{eq:optimal_velocity}
\end{equation}
The velocity is exactly the component of the post-selected state not yet acquired by the current state.
The factor $1/\braket{\chi}{\psi}$ in the weak vector is the same anomalous amplification familiar from weak-value physics; here it becomes the mathematical source of algorithmic amplification.
This is the dynamical content of the framework: the circuit is a way of realizing the weak-value drift dictated by the input and output boundary conditions.

Consequently, by explicitly calculating $H^*$ for specified initial and target boundary conditions, we can systematically uncover the computational procedure in the quantum algorithm required to traverse this optimal trajectory.

\section{Time-Invariance and Geometric Interpretation}
Before proceeding to specific applications, we highlight a critical property that simplifies the derivation of quantum algorithms: the time-invariance of the optimal Hamiltonian.

In general, the optimal Hamiltonian $H^*(t)$ derived in Eq.(\ref{eq:optimal_H}) depends explicitly on the time-evolving states $\ket{\psi(t)}$ and $\ket{\chi(t)}$. 
However, for an autonomous system where the cost function and constraints do not depend explicitly on time, the optimal control field is conserved \cite{Kirk2004, Pontryagin1962}. This is the optimal control analog of Noether's theorem (conservation of energy) \cite{Gelfand1963}.
Here, we provide a direct proof of this conservation law, $\frac{d}{dt}H^*(t) = 0$, which justifies the use of $t=0$ boundary conditions to determine the entire algorithm.
Let us differentiate the definition of $H^*(t)$ with respect to time:
\begin{equation}
    \frac{d}{dt} H^*(t) = \frac{i}{2} \left( \frac{d}{dt}(\ketbra{\chi}{\psi}) - \frac{d}{dt}(\ketbra{\psi}{\chi}) \right).
\end{equation}
The dynamics of the forward state $\ket{\psi}$ and the backward (adjoint) state $\ket{\chi}$ are governed by the same optimal Hamiltonian $H^*(t)$ (as derived from the canonical equations):
Substituting these equations of motion into the derivative, we apply the product rule:
\begin{align}
    \frac{d}{dt}(\ketbra{\psi}{\chi}) &= (-i H^*(t) \ket{\psi})\bra{\chi} + \ket{\psi}(\bra{\chi} i H^*(t)) \nonumber \\
    &= -i [H^*(t), \ketbra{\psi}{\chi}].
\end{align}
Similarly, for the second term, we have $\frac{d}{dt}(\ketbra{\chi}{\psi}) = -i [H^*(t), \ketbra{\chi}{\psi}]$.
Combining these results, the total time derivative becomes:
\begin{align}
    \frac{d}{dt} H^*(t) &= \frac{i}{2} \left( -i [H^*(t), \ketbra{\chi}{\psi}] + i [H^*(t), \ketbra{\psi}{\chi}] \right) \nonumber \\
    &= \frac{1}{2} [H^*(t), (\ketbra{\chi}{\psi} - \ketbra{\psi}{\chi})].
\end{align}
Noting that the term in the parentheses is proportional to $H^*(t)$ itself as in Eq. (\ref{eq:optimal_H}), the commutator vanishes.
This fact proves that the optimal Hamiltonian is strictly constant in time, $H^*(t) = H^*(0)$. 
Consequently, we can determine the global optimal control strategy solely by calculating Eq.(\ref{eq:optimal_H}) using the boundary conditions at $t=0$.
This seemingly counter-intuitive result---that a generator constructed from dynamical variables remains constant---exemplifies the power of the geometric approach: the optimal path is a geodesic, and the generator of a geodesic (the ``rotation axis") is an invariant of the motion.

The conservation law implies a constructive ``recipe" for algorithm discovery.
Instead of searching for a unitary circuit heuristically, one should identify the boundary operators $P_{in}$ and $P_{target}$ that stabilize (have as eigenstates) the initial and target states, respectively.
The optimal algorithmic generator is then essentially derived from their commutator:
\begin{equation}
    H^*_{algo} \propto i [P_{target}, P_{in}],
\end{equation}
up to the orientation convention of the path.
This algebraic relation explains why quantum algorithms are fundamentally built upon the non-commutativity of operations defined at the boundaries.
In the following sections, we demonstrate how this simple protocol automatically reconstructs the celebrated algorithms of Grover, Shor, and QSVT.

\subsection{Application I: Unstructured Search}
We now apply the QSB formalism to the unstructured search problem to derive Grover's algorithm. 
Let the search space size be $N=2^n$. We define the initial state as the uniform superposition $\ket{s} = \sum_x \ket{x}/\sqrt{N}$ and the target state as the solution $\ket{w}$. 
We assume the overlap $\sin\theta \equiv \braket{w}{s} = 1/\sqrt{N}$ is small but non-zero.

We seek the optimal control Hamiltonian $H^*$ that transports $\ket{s}$ to $\ket{w}$ with minimum energy. 
Substituting $\ket{\psi}=\ket{s}$ and $\ket{\chi}=\ket{w}$ into the general solution Eq.(\ref{eq:optimal_H}), we obtain the instantaneous optimal Hamiltonian:
\begin{equation}
    H^*_{Grover} = \frac{i}{2} \left( \ketbra{w}{s} - \ketbra{s}{w} \right).
    \label{eq:Grover_QSB_raw}
\end{equation}
To elucidate the physical action of this Hamiltonian, we introduce the Gram-Schmidt orthogonalization with respect to the target state.
Let $\ket{w^\perp}$ be the normalized state component of $\ket{s}$ that is orthogonal to $\ket{w}$, defined such that $\ket{s} = \sin\theta \ket{w} + \cos\theta \ket{w^\perp}$.
Substituting this expansion into Eq.(\ref{eq:Grover_QSB_raw}), the terms involving $\ketbra{w}{w}$ cancel out, leaving:
\begin{equation}
    H^*_{Grover} = \frac{i}{2} \cos\theta \left( \ketbra{w}{w^\perp} - \ketbra{w^\perp}{w} \right).
    \label{eq:Grover_Generator}
\end{equation}
This operator is the generator of a rotation in the two-dimensional subspace $\mathcal{H}_{2D} = \text{span}\{\ket{w^\perp}, \ket{w}\}$. 
Specifically, in the basis $\{\ket{w^\perp}, \ket{w}\}$, it corresponds to the Pauli matrix $\sigma_y$ (up to a prefactor), generating the unitary evolution $U(t) = \exp(-i H^*_{Grover} t)$.
This evolution rotates the state vector directly from the initial non-solution component $\ket{w^\perp}$ towards the target $\ket{w}$ along the great circle (geodesic).

The crucial step is to realize this optimal continuous rotation using accessible quantum gates. 
In the standard circuit model, we cannot directly implement interactions involving $\ket{w^\perp}$ as we do not know $\ket{w}$ a priori. 
However, we have access to two specific operations: the ``oracle" reflection $R_w = I - 2\ketbra{w}{w}$ and the ``diffusion" reflection $R_s = 2\ketbra{s}{s} - I$.

Geometrically, the product of two reflections in a plane is equivalent to a rotation by twice the angle between the reflection axes. 
This geometry has a direct algebraic counterpart in quantum mechanics, isomorphic to the theory of angular momentum (or spin-1/2 systems).
Let us verify the connection to $H^*_{Grover}$. 
Consider the commutator of the reflection operators $R_w$ and $R_s$:
\begin{align}
    [R_w, R_s] &= 8i \sin\theta H^*_{Grover}.
\end{align}
This result is profound: it mirrors the fundamental commutation relation of the Pauli matrices, $[\sigma_z, \sigma_x] = 2i\sigma_y$.
Just as non-commuting rotations around the $z$ and $x$ axes generate a rotation around the $y$ axis, the non-commutativity of the oracle and diffusion reflections generates the optimal Hamiltonian $H^*_{Grover}$.
The discrete Grover iteration $G = R_s R_w$ is the finite-reflection realization of the same Lie algebra: the product of two reflections is a rotation in $\mathcal{H}_{2D}$, and the rotation axis is precisely the QSB generator.
Equivalently, at the projector level,
\begin{equation}
    [P_w,P_s] = -2i\sin\theta\, H^*_{Grover},
\end{equation}
where $P_s=\ketbra{s}{s}$ and $P_w=\ketbra{w}{w}$.
A Baker-Campbell-Hausdorff expansion of short pulses generated by $P_s$ and $P_w$ exposes the same commutator direction, while the usual Grover reflections implement its finite-angle version exactly in the two-dimensional invariant subspace.
Consequently, the oracle and diffusion operators emerge not as heuristic gadgets, but as the necessary boundary reflections required to synthesize the optimal weak-value drift derived from the QSB variational principle.

\subsection{Application II: Structured Search}
The power of the QSB framework is most evident when the target state possesses a specific algebraic structure, as seen in Shor's factorization algorithm. 
The core of Shor's quantum advantage lies in the order finding subroutine, which relies on the quantum Fourier transform (QFT) to extract periodicity. 
Thus, deriving Shor's algorithm via QSB reduces to finding the optimal control that generates the QFT.

We define the problem as transporting a computational basis state $\ket{x} = \ket{x_1 x_2 \dots x_n}$ (where $x = \sum x_j 2^{n-j}$) to its Fourier transform, up to the conventional final bit reversal.
The target state is explicitly given by the product state:
\begin{equation}
    \ket{\psi_{target}} = \text{QFT}\ket{x} = \frac{1}{\sqrt{N}} \bigotimes_{k=1}^n \left( \ket{0} + e^{i \phi_k(x)} \ket{1} \right)_k,
\end{equation}
where the phase $\phi_k(x)$ for the $k$-th qubit is determined by the binary expansion of the input $x$:
\begin{equation}
    \phi_k(x) = 2\pi (0.x_k x_{k+1} \dots x_n) = 2\pi \sum_{j=k}^n \frac{x_j}{2^{j-k+1}}.
\end{equation}
Unlike Grover's problem, here the target state is a product state. 
This structure implies that the optimal transport cost is additive, and the generator decouples into each qubit $k$. 

It is useful to separate the QFT into two boundary tasks.
The local Hadamard layer creates an equatorial state; when $x_k=1$, the extra $\pi$ phase is precisely the first term in $\phi_k(x)$.
After absorbing this bit-dependent phase into $\phi_k(x)$, the remaining phase-imprinting boundary can be written as $\ket{+}\mapsto(\ket{0}+e^{i\phi_k(x)}\ket{1})/\sqrt{2}$.
We therefore derive the phase-imprinting generator $H^{(k)}$ from this single-qubit boundary condition.
The initial equatorial state has the projector $P_{in} = \ketbra{+}{+} = \frac{1}{2}(I + X_k)$.
The target state for the $k$-th qubit is $\ket{\chi} = \frac{1}{\sqrt{2}}(\ket{0} + e^{i\phi_k(x)}\ket{1})$, which corresponds to a state rotated by angle $\phi_k$ around the Z-axis. Its projector is given by:
\begin{equation}
    P_{target} = \frac{1}{2} \left( I + \cos\phi_k X_k + \sin\phi_k Y_k \right).
\end{equation}

By choosing the global phase of the target state to minimize the geodesic distance (ensuring real overlap), the optimal Hamiltonian Eq.(\ref{eq:optimal_H}) becomes proportional to the commutator of these projectors, $H^{(k)} \propto i [P_{target}, P_{in}]$. 
Let us calculate this commutator explicitly:
\begin{align}
    [P_{target}, P_{in}]
    &= \frac{\sin\phi_k}{4} [Y_k, X_k] = -\frac{i}{2} \sin\phi_k Z_k.
\end{align}
Multiplying by $i$ (from the QSB formula), we find that the optimal generator is strictly proportional to the Pauli-Z operator:
\begin{equation}
    H^{(k)} \propto Z_k.
\end{equation}
This derivation proves that the optimal control field is necessarily diagonal in the computational basis. 
Since $Z_k = \ketbra{0}{0} - \ketbra{1}{1} = 1 - 2\ketbra{1}{1}$, and the identity term contributes only a global phase, the physical generator is the number operator $\hat{n}_k = \ketbra{1}{1}_k$.
The strength of the field is determined by the total rotation angle required, $\phi_k(x)$. Thus, the Hamiltonian is uniquely determined as:
\begin{equation}
    H^{(k)}(x) = -\frac{\phi_k(x)}{T} \ketbra{1}{1}_k,
\end{equation}
where the sign corresponds to the conventional QFT phase choice in the target state; reversing it gives the inverse QFT.

To construct the quantum algorithm, we promote the classical parameter $x_j$ in $\phi_k(x)$ to a quantum operator acting on the input register. 
We replace the classical bit $x_j$ with the number operator $\hat{n}_j = \ketbra{1}{1}_j$.
Substituting the expression for $\phi_k(x)$, the total Hamiltonian $H^*_{Shor} = \sum_{k=1}^n H^{(k)}(\hat{n})$ becomes:
\begin{align}
    H^*_{Shor} 
    &= -\frac{2\pi}{T} \sum_{k=1}^n \left( \frac{\hat{n}_k}{2} + \sum_{j=k+1}^n \frac{\hat{n}_j \hat{n}_k}{2^{j-k+1}} \right).
    \label{eq:Shor_Hamiltonian}
\end{align}
Here, using the projector property $\hat{n}_k^2 = \hat{n}_k$, we can separate the Hamiltonian into single-body and two-body terms.

Equation (\ref{eq:Shor_Hamiltonian}) represents the exact Hamiltonian that generates the Quantum Fourier Transform. Since all terms involve only number operators $\hat{n}_k$ (or effectively $Z_k$), they commute with each other. Thus, the time-evolution operator can be factorized exactly without Trotter error:
\begin{equation}
    U(T) = \left( \prod_{k=1}^n e^{i \pi \hat{n}_k} \right) \left( \prod_{k < j} e^{i \frac{\pi}{2^{j-k}} \hat{n}_j \hat{n}_k} \right).
\end{equation}

This factorization maps directly onto the standard quantum circuit components.
The term $e^{i \pi \hat{n}_k}$ corresponds to a phase gate $R_1$ (a Z-rotation by $\pi$) on the $k$-th qubit. In the standard QFT decomposition, this is often absorbed into the Hadamard gate sequence or represents the fundamental phase resolution.
On the other hand, the term $\exp\left( i \pi \hat{n}_j \hat{n}_k/2^{j-k}  \right)$ acts non-trivially only when both qubits $j$ and $k$ are in the state $\ket{1}$. 
This is precisely the definition of the controlled-phase rotation gate $CR_m(\theta)$ with angle $\theta = \pi / 2^{m}$, where $m = j-k$ is the ``distance" between the qubits.

While it is well-established that the QFT circuit consists of these gates \cite{Coppersmith1994}, our derivation provides a novel physical insight: the two-body Ising-like interactions arise necessarily as the optimal control solution to satisfy the phase correlations inherent in the target distribution. 
In the QSB framework, the computational ``speedup" of Shor's algorithm can be attributed to the efficient decomposition of the weak value drift into local pairwise interactions. 
The algorithm is not an arbitrary sequence of operations but the natural physical evolution of a system tuned to minimize the transport cost to the Fourier basis.

\subsection{Application III: Polynomial Transformation}
Finally, we address the quantum singular value transformation (QSVT), which unifies a vast array of quantum algorithms, from Hamiltonian simulation to matrix inversion \cite{Gilyen2019, Martyn2021}.
In the QSB framework, QSVT is naturally interpreted as a singular-value-resolved version of Grover rotation.
Grover's algorithm rotates one two-dimensional search subspace by a fixed angle; QSVT rotates every singular-value subspace by a prescribed angle that encodes an admissible polynomial transformation.

Let $A$ be a contraction with singular-value decomposition $A=\sum_k \sigma_k \ketbra{u_k}{v_k}$, block-encoded in a unitary $U_A$.
QSVT does not implement an arbitrary function.
For a phase sequence of length $d$, the real polynomial $p$ appearing in the transformed block must satisfy the standard QSP/QSVT admissibility conditions \cite{Low2017,Gilyen2019}: $\deg p\le d$, $p$ has the parity fixed by $d$, $|p(x)|\le 1$ for $x\in[-1,1]$, and $p$ admits a unitary completion, equivalently there exists a polynomial $q$ such that
\begin{equation}
    |p(x)|^2 + (1-x^2)|q(x)|^2 = 1
    \qquad (x\in[-1,1]).
    \label{eq:qsvt_completion}
\end{equation}
These conditions have a direct bridge interpretation.
The bound and completion condition guarantee that the target boundary is normalizable in each invariant signal subspace, while the parity condition characterizes the reachable boundary profiles generated by alternating the signal unitary with single-qubit phase rotations.

For each singular value $\sigma_k$, the block encoding defines an invariant two-dimensional signal subspace $\mathcal{H}_k$.
In that subspace we may write the desired boundary condition as
\begin{equation}
    \ket{\psi_{in}^{(k)}} \mapsto
    \ket{\psi_{target}^{(k)}} =
    p(\sigma_k)\ket{0_k}
    + \sqrt{1-|p(\sigma_k)|^2}\ket{1_k},
\end{equation}
where $\{\ket{0_k},\ket{1_k}\}$ denotes an orthonormal basis of $\mathcal{H}_k$ fixed by the block encoding and the chosen unitary completion.
Applying the QSB formula Eq.(\ref{eq:optimal_H}) to this boundary pair gives the continuous optimal generator
\begin{equation}
    H^*_{p} = \sum_k \frac{\Theta(\sigma_k)}{T} \sigma_y^{(k)},
    \qquad
    \Theta(\sigma_k)=\arccos p(\sigma_k),
\end{equation}
where $\sigma_y^{(k)}$ is the Pauli-Y operator defined in the effective qubit subspace $\mathcal{H}_k$.
Thus the weak-value drift is now a family of drifts whose strength depends nonlinearly on the singular value through the target polynomial.

We cannot directly apply $H^*_{p}$ as a continuous singular-value-dependent Hamiltonian.
However, we have access to two operations that define the ``axes" of rotation.
In the subspace $\mathcal{H}_k$, the block encoding $U_A$ acts as a signal-dependent rotation, with angle determined by $\sigma_k$.
The projector-controlled phase rotation
\begin{equation}
    U(\phi)=e^{-i\phi(2\ketbra{0}{0}-I)\otimes I}
\end{equation}
provides the adjustable control axis.

Just as in Grover's algorithm where the commutator of the boundary reflections generated the drift, here the algebraic interplay between $U_A$ and $U(\phi)$ generates the desired singular-value-dependent drift.
By applying an alternating sequence of these operators,
\begin{equation}
    U_{\vec{\phi}} = \prod_j U(\phi_j)U_A,
\end{equation}
one synthesizes an element of the same $\mathrm{SU}(2)$ Lie algebra in every $\mathcal{H}_k$.
The QSP/QSVT existence theorem states precisely that, for every admissible $p$, there is a phase sequence $\vec{\phi}$ such that the block of $U_{\vec{\phi}}$ is $p(A)$, equivalently the induced $\mathrm{SU}(2)$ evolution has matrix element $p(\sigma_k)$ for all $k$ simultaneously \cite{Low2017,Gilyen2019}.

Thus, the QSB perspective clarifies the physics of QSVT:
The signal operator $U_A$ provides the raw connection to the singular values (the ``oracle" equivalent), and the signal processing operator  $U(\phi)$ provides the adjustable control (the ``diffusion" equivalent).
The sequence of phases $\vec{\phi}$ is the discretized control protocol required to synthesize the QSB boundary profile from the available non-commuting generators.
The parity, boundedness, and completion constraints are not technical afterthoughts; they are the exact controllability conditions for realizing a unitary bridge simultaneously across all singular-value subspaces.

The preceding examples have exact or algebraically specified output boundaries.
Variational algorithms such as QAOA \cite{Farhi2014} lie outside this regime, because the desired ground-state boundary is not known in advance.
In such cases the QSB generator $i[P_{target},P_{in}]$ cannot be computed directly; one may only approximate the unknown bridge by using a cost Hamiltonian, a mixer, and variationally chosen controls.
We therefore regard QAOA-like constructions not as further evidence for the exact results above, but as a natural direction for approximate QSB design.
A conservative strategy is to introduce intermediate boundaries with appreciable pairwise overlap and to apply the local bridge construction segment by segment.
This ``iterative bridge" viewpoint is a proposal for future work.

\section{Conclusion}
We have presented the coherent Quantum Schr\"odinger Bridge as a two-boundary principle for quantum algorithm design.
The input state and the desired output state define a pre- and post-selected ensemble; Aharonov's two-state vector supplies the kinematics of this ensemble, weak values supply its local response, and optimal control converts that response into a Hamiltonian flow.
This is the conceptual link: a quantum algorithm is a physical procedure for realizing the imaginary weak-value drift from the input boundary to the output boundary.

Grover search, the QFT, and QSVT then appear as different realizations of the same mechanism.
Grover's oracle and diffusion reflections synthesize the drift between two boundary projectors.
The QFT controlled phases synthesize the drift required to write global arithmetic information into relative phases.
QSVT synthesizes an entire family of singular-value-dependent drifts by alternating the block encoding with controllable phase rotations.
The common structure is not the surface form of the circuit, but the Lie algebra generated by non-commuting controls and computational boundary conditions.

This perspective also suggests why variational algorithms are difficult in high-dimensional spaces \cite{McClean2018}.
When the final boundary is unknown or nearly orthogonal to the initial ansatz, the useful weak-value response is hard to access with the available controls.
Warm starts, problem-inspired ansatzes, and intermediate-state constructions may be understood as attempts to restore boundary pairs with appreciable local response.

Thus the message of the present work is not that every quantum algorithm is already known once the target state is named.
Rather, once the computational input and output boundaries are specified together with the available controls, QSB provides the variational object that the circuit must synthesize.
It turns weak measurement, optimal control, and quantum algorithm design into a single language for discovering and organizing quantum circuits.

\section*{Acknowledgement}
We received financial support from the Cross-ministerial Strategic Innovation Promotion Program (SIP) from the Cabinet Office (No. 23836436).
\bibliography{references}

@article{Grover1997,
  title = {Quantum Mechanics Helps in Searching for a Needle in a Haystack},
  author = {Grover, Lov K.},
  journal = {Phys. Rev. Lett.},
  volume = {79},
  issue = {2},
  pages = {325--328},
  numpages = {4},
  year = {1997},
  publisher = {American Physical Society},
  doi = {10.1103/PhysRevLett.79.325}
}

@inproceedings{Shor1994,
  title={Algorithms for quantum computation: discrete logarithms and factoring},
  author={Shor, Peter W.},
  booktitle={Proceedings of the 35th Annual Symposium on Foundations of Computer Science (FOCS)},
  pages={124--134},
  year={1994},
  organization={IEEE},
  doi={10.1109/SFCS.1994.365700}
}

@article{Coppersmith1994,
  title={An approximate Fourier transform useful in quantum factoring},
  author={Coppersmith, Don},
  journal={arXiv preprint quant-ph/0201067},
  year={1994},
  note={IBM Research Report RC19642}
}

@article{Carlini2006,
  title = {Time-Optimal Quantum Evolution},
  author = {Carlini, Alberto and Hosoya, Akio and Koike, Tatsuhiko and Okudaira, Yosuke},
  journal = {Phys. Rev. Lett.},
  volume = {96},
  issue = {6},
  pages = {060503},
  numpages = {4},
  year = {2006},
  publisher = {American Physical Society},
  doi = {10.1103/PhysRevLett.96.060503}
}

@article{Schrodinger1931,
  title={{\"U}ber die Umkehrung der Naturgesetze},
  author={Schr{\"o}dinger, Erwin},
  journal={Sitzungsberichte der Preussischen Akademie der Wissenschaften. Physikalisch-mathematische Klasse},
  volume={144},
  pages={144--153},
  year={1931}
}

@article{GeorgiouPavon2015,
  title={Positive contraction mappings for classical and quantum Schr\"odinger systems},
  author={Georgiou, Tryphon T. and Pavon, Michele},
  journal={Journal of Mathematical Physics},
  volume={56},
  number={3},
  pages={033301},
  year={2015},
  publisher={AIP Publishing},
  doi={10.1063/1.4915289}
}

@article{BeckerLi2021,
  title={Quantum Statistical Learning via Quantum Wasserstein Natural Gradient},
  author={Becker, Simon and Li, Wuchen},
  journal={Journal of Statistical Physics},
  volume={182},
  number={1},
  pages={1--26},
  year={2021},
  publisher={Springer},
  doi={10.1007/s10955-020-02682-1}
}

@article{Movilla2025,
  title={Quantum Schr\"odinger bridges: large deviations and time-symmetric ensembles},
  author={Movilla Miangolarra, Olga and Sabbagh, Ralph and Georgiou, Tryphon T.},
  journal={arXiv preprint arXiv:2503.05886},
  year={2025}
}

@article{Khaneja2001,
  title = {Time Optimal Control in Spin Systems},
  author = {Khaneja, Navin and Brockett, Roger and Glaser, Steffen J.},
  journal = {Phys. Rev. A},
  volume = {63},
  pages = {032308},
  year = {2001},
  publisher = {American Physical Society},
  doi = {10.1103/PhysRevA.63.032308}
}

@article{Aharonov1964,
  title = {Time Symmetry in the Quantum Process of Measurement},
  author = {Aharonov, Yakir and Bergmann, Peter G. and Lebowitz, Joel L.},
  journal = {Phys. Rev.},
  volume = {134},
  issue = {6B},
  pages = {B1410--B1416},
  year = {1964},
  publisher = {American Physical Society},
  doi = {10.1103/PhysRev.134.B1410}
}

@article{AharonovVaidman1991,
  title = {Complete description of a quantum system at a given time},
  author = {Aharonov, Yakir and Vaidman, Lev},
  journal = {Journal of Physics A: Mathematical and General},
  volume = {24},
  number = {10},
  pages = {2315--2328},
  year = {1991},
  doi = {10.1088/0305-4470/24/10/018}
}

@article{Aharonov1988,
  title = {How the result of a measurement of a component of the spin of a spin-1/2 particle can turn out to be 100},
  author = {Aharonov, Yakir and Albert, David Z. and Vaidman, Lev},
  journal = {Phys. Rev. Lett.},
  volume = {60},
  issue = {14},
  pages = {1351--1354},
  year = {1988},
  publisher = {American Physical Society},
  doi = {10.1103/PhysRevLett.60.1351}
}

@article{Nielsen2006,
  title = {Quantum Computation as Geometry},
  author = {Nielsen, Michael A. and Dowling, Mark R. and Gu, Mile and Doherty, Andrew C.},
  journal = {Science},
  volume = {311},
  number = {5764},
  pages = {1133--1135},
  year = {2006},
  publisher = {American Association for the Advancement of Science},
  doi = {10.1126/science.1121541},
  note = {Establishes the minimization of Hamiltonian norms as finding geodesics on the unitary manifold.}
}

@article{Peirce1988,
  title = {Optimal control of quantum-mechanical systems: Existence, numerical approximation, and applications},
  author = {Peirce, A. P. and Dahleh, M. A. and Rabitz, H.},
  journal = {Phys. Rev. A},
  volume = {37},
  issue = {12},
  pages = {4950--4964},
  year = {1988},
  publisher = {American Physical Society},
  doi = {10.1103/PhysRevA.37.4950},
  note = {Foundational work on energy-optimal quantum control ($L^2$ minimization).}
}

@article{Dressel2014,
  title = {Colloquium: Understanding quantum weak values: Basics and applications},
  author = {Dressel, Justin and Malik, Mehul and Miatto, Filippo M. and Jordan, Andrew N. and Boyd, Robert W.},
  journal = {Rev. Mod. Phys.},
  volume = {86},
  issue = {1},
  pages = {307--316},
  year = {2014},
  publisher = {American Physical Society},
  doi = {10.1103/RevModPhys.86.307},
  note = {Comprehensive review detailing how the imaginary part of the weak value relates to back-action and drift in the conjugate variable.}
}

@book{Kirk2004,
  title={Optimal Control Theory: An Introduction},
  author={Kirk, Donald E.},
  year={2004},
  publisher={Dover Publications},
  address={New York},
  note={See Chapter 5.2 for the proof that the Hamiltonian is constant for time-invariant systems.}
}

@article{Low2017,
  title={Optimal Hamiltonian Simulation by Quantum Signal Processing},
  author={Low, Guang Hao and Chuang, Isaac L.},
  journal={Physical Review Letters},
  volume={118},
  number={1},
  pages={010501},
  year={2017},
  publisher={American Physical Society},
  doi={10.1103/PhysRevLett.118.010501}
}

@inproceedings{Gilyen2019,
  title={Quantum singular value transformation and beyond: exponential improvements for quantum matrix arithmetics},
  author={Gily{\'e}n, Andr{\'a}s and Su, Yuan and Low, Guang Hao and Wiebe, Nathan},
  booktitle={Proceedings of the 51st Annual ACM SIGACT Symposium on Theory of Computing (STOC)},
  pages={193--204},
  year={2019},
  organization={ACM},
  doi={10.1145/3313276.3316366}
}

@article{Farhi2014,
  title={A Quantum Approximate Optimization Algorithm},
  author={Farhi, Edward and Goldstone, Jeffrey and Gutmann, Sam},
  journal={arXiv preprint arXiv:1411.4028},
  year={2014}
}

@article{McClean2018,
  title={Barren plateaus in quantum neural network training landscapes},
  author={McClean, Jarrod R. and Boixo, Sergio and Smelyanskiy, Vadim N. and Babbush, Ryan and Neven, Hartmut},
  journal={Nature Communications},
  volume={9},
  number={1},
  pages={4812},
  year={2018},
  publisher={Nature Publishing Group}
}

@article{Martyn2021,
  title={Grand Unification of Quantum Algorithms},
  author={Martyn, John M. and Rossi, Zane M. and Tan, Andrew K. and Chuang, Isaac L.},
  journal={PRX Quantum},
  volume={2},
  number={4},
  pages={040203},
  year={2021},
  publisher={American Physical Society},
  doi={10.1103/PRXQuantum.2.040203}
}

@book{Gelfand1963,
  title={Calculus of Variations},
  author={Gelfand, I. M. and Fomin, S. V.},
  year={1963},
  publisher={Dover Publications},
  address={New York},
  note={Classic text discussing the connection between time-invariance and conservation of energy (Noether's theorem) in variational problems.}
}

@book{Pontryagin1962,
  title={The Mathematical Theory of Optimal Processes},
  author={Pontryagin, L.S. and Boltyanskii, V.G. and Gamkrelidze, R.V. and Mishchenko, E.F.},
  year={1962},
  publisher={Interscience},
  address={New York},
  note={The foundational text on Pontryagin's Minimum Principle.}
}

\end{document}